\documentclass[aps,prd,amssymb,superscriptaddress,floatfix,nofootinbib,10pt]{revtex4}
\usepackage{times,amsbsy,amsmath,amsfonts,graphicx,float}
\usepackage{color,morefloats,rotating,srcltx,slashed}
\usepackage{multirow,bm,verbatim,tabularx,bbding,threeparttable}
\definecolor{dblue}{rgb}{0.00,0.00,0.75}
\usepackage[colorlinks,urlcolor=dblue,linkcolor=dblue,citecolor=dblue]{hyperref} 
\allowdisplaybreaks[4]

\newcommand{\PreserveBackslash}[1]{\let\temp=\\#1\let\\=\temp}
\newcolumntype{C}[1]{>{\PreserveBackslash\centering}p{#1}}
\newcolumntype{R}[1]{>{\PreserveBackslash\raggedleft}p{#1}}
\newcolumntype{L}[1]{>{\PreserveBackslash\raggedright}p{#1}}

\begin{document}

                            \title{Molecular $\Omega_{cc}$ , $\Omega_{bb}$ and $\Omega_{bc}$ states}

\author{W.~F. Wang}
	\email{wfwang@ific.uv.es}
	\affiliation{Institute of Theoretical Physics, Shanxi University, Taiyuan, Shanxi 030006, China}
	\affiliation{Departamento de F\'{\i}sica Te\'orica and IFIC, Centro Mixto Universidad de
	                 Valencia-CSIC Institutos de Investigaci\'on de Paterna, Aptdo.~22085, 46071 Valencia, Spain}
                 
\author{A. Feijoo}
	\email{edfeijoo@ific.uv.es}
	\affiliation{Departamento de F\'{\i}sica Te\'orica and IFIC, Centro Mixto Universidad de
	                 Valencia-CSIC Institutos de Investigaci\'on de Paterna, Aptdo.~22085, 46071 Valencia, Spain}

\author{J. Song}
	\email{song-jing@buaa.edu.cn}
	\affiliation{School of space and environment, Beihang University, Beijing, 102206, China}
	\affiliation{School of Physics, Beihang University, Beijing, 102206, China}
	\affiliation{Departamento de F\'{\i}sica Te\'orica and IFIC, Centro Mixto Universidad de
	                 Valencia-CSIC Institutos de Investigaci\'on de Paterna, Aptdo.~22085, 46071 Valencia, Spain}

\author{E. Oset}
        \thanks{Corresponding author}
	\email{oset@ific.uv.es}
	\affiliation{Departamento de F\'{\i}sica Te\'orica and IFIC, Centro Mixto Universidad de
       		          Valencia-CSIC Institutos de Investigaci\'on de Paterna, Aptdo.~22085, 46071 Valencia, Spain}

\begin{abstract}
We study the interaction of meson-baryon coupled channels carrying quantum numbers of $\Omega_{cc}$ , $\Omega_{bb}$ and $\Omega_{bc}$ presently under investigation by the LHCb collaboration. The interaction is obtained from an extension of the local hidden gauge approach to the heavy quark sector that has proved to provide accurate results compared to experiment in the case of $\Omega_{c}$ , $\Xi_{c}$ states and pentaquarks, $P_c$ and $P_{cs}$. We obtain many bound states, with small decay widths within the space of the chosen coupled channels. The spin-parity of the states are $J^P={\frac{1}{2}}^-$ for coupled channels of pseudoscalar-baryon (${\frac{1}{2}}^+$), $J^P={\frac{3}{2}}^-$ for the case of pseudoscalar-baryon (${\frac{3}{2}}^+$), $J^P={\frac{1}{2}}^-,{\frac{3}{2}}^-$ for the case of vector-baryon (${\frac{1}{2}}^+$) and $J^P={\frac{1}{2}}^-,{\frac{3}{2}}^-,{\frac{5}{2}}^-$ for the vector-baryon (${\frac{3}{2}}^+$) channels. We look for poles of the states and evaluate the couplings to the different channels. The couplings obtained for the open channels can serve as a guide to see in which reaction the obtained states are more likely to be observed.
\end{abstract}

\date{\today}

\maketitle

\section{Introduction}
The combined efforts in present hadron facilities are giving rise to the discovery of many new states with heavy quarks, some of them manifestly exotic, which do not follow the standard rules in terms of quarks of $q \bar q$ for mesons and $qqq$ for baryons. In particular, observations of baryons $\Lambda_c$~\cite{CLEO:1994oxm,ARGUS:1993vtm,CLEO:2000mbh,LHCb:2017jym,BaBar:2006itc} , $\Sigma_c$~\cite{Belle:2004zjl} , $\Xi_c$~\cite{CLEO:2000ibb,CLEO:1999msf,BaBar:2007xtc,LHCb:2020iby,Belle:2006edu,Belle:2020tom,BaBar:2007zjt} , $\Omega_c$~\cite{LHCb:2017uwr}, $\Lambda_b$~\cite{LHCb:2012kxf,CDF:2013pvu,CMS:2020zzv,LHCb:2020lzx,LHCb:2019soc}, $\Sigma_b$~\cite{LHCb:2018haf} , $\Xi_b$~\cite{CMS:2021rvl,LHCb:2018vuc,LHCb:2020xpu,LHCb:2021ssn}, $\Omega_b$~\cite{LHCb:2020tqd}  have been reported (see Ref.~\cite{Chen:2022asf}  for a recent review of experimental findings). The discovery of the hidden charm $P_c$ pentaquarks~\cite{LHCb:2015yax,LHCb:2019kea,LHCb:2021chn}  and hidden charm with strangeness $P_{cs}$~\cite{LHCb:2020jpq} has added extra excitement to the field.  The search for new states continues, and motivated by the LHCb plans to measure new states, we concentrate here in the theoretical study of the $\Omega_{cc}$, $\Omega_{bb}$ and $\Omega_{bc}$ which are presently under investigation by the LHCb collaboration\footnote{M. Pappagallo in a talk at the SNP School 2021, http://lambda.phys.tohoku.ac.jp/snpsc2021/}. We should note that some attempts to search for $\Omega_{bc}$  and $\Xi_{bc}$ have already been conducted, so far with inconclusive results~\cite{LHCb:2021xba}.

These states have been the subject of intense investigation in the past using quark models~\cite{Gershtein:1998sx,Gershtein:2000nx,Ebert:2002ig,Kiselev:2002iy,Roberts:2007ni,Valcarce:2008dr,Albertus:2009ww,Albertus:2012jt,Ma:2015lba,Shah:2016vmd,Garcilazo:2016piq,Lu:2017meb,Xiao:2017dly,Niu:2018ycb,Salehi:2018oky,Li:2019ekr,Gutierrez-Guerrero:2019uwa,Shah:2021reh,Ghalenovi:2022dok}. They have also been studied in the framework of QCD lattice~\cite{Garcilazo:2016piq,Lewis:2001iz,Lin:2011ti,Brown:2014ena,Perez-Rubio:2015zqb,Bahtiyar:2018vub,Briceno:2012wt} and the framework of QCD sum rules~~\cite{Wang:2010hs,Hu:2017dzi,Wang:2018lhz}. Also, different approaches have been followed in~\cite{Weng:2010rb,Karliner:2015ina,Soto:2020pfa,Soto:2021cgk}.

Our aim is to study the states of this type that can be formed as molecular states from the interaction in s-wave of mesons with baryons in their ground state. Hence, one anticipates that we shall only obtain baryon states with negative parity, different to most of the states obtained from quark models. The attractive force between mesons and baryons in many cases makes the appearance of these states unavoidable, as has been discussed in detail in~\cite{Dong:2021bvy,Dong:2021juy}. Also, the proximity of the mass of some states to the threshold of some meson baryon channel introduces constraints that require the explicit consideration of these channels and their interaction in a study of the baryon spectrum~\cite{Dong:2021rpi}.

The work in the molecular field of meson-baryon interaction is vast and is reviewed in~\cite{Oller:2000ma,Guo:2017jvc}. Concerning baryons that contain heavy quarks, work has been done by different groups. The information of chiral Lagrangians is extrapolated to the charm sector to study the $\Lambda_c(2595)$ using the $DN$ and $\Sigma_c \pi$ coupled channels in~\cite{Hofmann:2005sw,Mizutani:2006vq,Tolos:2007vh,Garcia-Recio:2008rjt}. The molecular $\Omega_c$ states were studied  in~\cite{Hofmann:2005sw,Jimenez-Tejero:2009cyn,Romanets:2012hm,Montana:2017kjw}. A similar approach was used in~\cite{Debastiani:2017ewu}, differing in the use of baryon wave functions, which were borrowed from~\cite{Roberts:2007ni} and made unnecessary to invoke elements of SU$(4)$ symmetry used in former works. The work of~\cite{Debastiani:2017ewu} uses an extrapolation to the charm sector of the local hidden gauge approach of~\cite{Bando:1984ej,bando1988nonlinear,Harada:2003jx,Meissner:1987ge,Nagahiro:2008cv} and was successful to reproduce three of the $\Omega_c$ states reported in~\cite{LHCb:2017uwr}. The successful scheme, exchanging vector mesons, has then been used to make predictions for several types of baryonic states containing open charm or bottom. In this sense in~\cite{Dias:2018qhp}  the $\Xi_{cc}$ states have been studied, in~\cite{Dias:2019klk} that work is extended to study $\Xi_{bb}$ and $\Omega_{bbb}$ states, in~\cite{Yu:2018yxl}  to study the $\Xi_c$ and $\Xi_b$ states, in~\cite{Yu:2019yfr} to study the $\Xi_{bc}$ states and in ~\cite{Liang:2017ejq} to study $\Omega_b$ states. Some of these states can be associated to experimental states recently found~\cite{Chen:2022asf}  and in the case of $\Omega_b$ it is shown in~\cite{Liang:2020dxr}  that the states predicted in~\cite{Liang:2017ejq}  could have already been observed in the experiment of Ref.~\cite{LHCb:2020tqd}  in the higher part of the spectrum.  Each of the cases requires an elaborate study of the interaction with many coupled channels.

Other works also look at these states from the molecular point of view using different formalism and dynamics. In~\cite{Shimizu:2017xrg}  one pion exchange and $D^{(*)}$ exchange are used as a source of the dynamics to study $\Xi_{cc}$ states. In~\cite{Guo:2013xga}, heavy flavor, heavy quark spin, and heavy antiquark-diquark symmetries for hadronic molecules are considered within an effective field theory framework to study pentaquarks and open bottom, baryonic states. In~\cite{Chen:2017xat}  the one boson exchange picture is used to study $\Omega_c$ states. Particular mention deserves the extended work in different sectors, light, charm and bottom of molecular states stemming from meson baryon interaction in coupled channels using  $SU(6)_{\mathrm{lsf}} \times SU(2)\mathrm{HQSS}$ symmetry, this is $SU(6)$ flavor-spin symmetry in the light sector and $SU(2)$ in the heavy sector respecting heavy quark spin symmetry~\cite{Romanets:2012hm,Garcia-Recio:2013gaa,Garcia-Recio:2012lts}, extrapolating dynamics from the  Weinberg Tomozawa interaction. These works have the virtue of correlating many coupled channels in different sectors and make qualitative predictions for bound states and resonances in a large span of quantum numbers. $\Xi_b$ and  $\Xi_c$ states have been recenty addressed from this point of view in~\cite{Nieves:2019jhp}.

A large number of studies of molecular states in the charm sector have been devoted to the study of pentaquarks $P_c$ and $P_{cs}$. These states have hidden charm and we are concerned about open charm and bottom, so we refrain from discussing this issue and address the readers to review papers that also serve as reviews for open heavy quark baryonic molecular states~\cite{Dong:2021bvy,Dong:2021juy,Guo:2017jvc,Chen:2016qju,Dong:2017gaw,Liu:2019zoy,Ramos:2020bgs}.

\section{Formalism}

\subsection{Coupled channels for the $\Omega_{cc}$, $\Omega_{bb}$ and $\Omega_{bc}$ states}

In the first place, we must select the coupled channels that we consider in the approach. The task is facilitated by looking at the work of Ref~\cite{Romanets:2012hm}. We choose the same coupled channels eliminating only a few that appear at too high energy compared to the bulk of them. After that, the coupled channels are separated into four blocks, $PB({\frac{1}{2}}^+)$, $PB({\frac{3}{2}}^+)$, $VB({\frac{1}{2}}^+)$, $VB({\frac{3}{2}}^+)$, where $P$ stands for the pseudoscalar meson, $V$ for vector meson, $B({\frac{1}{2}}^+)$ for ground state baryons with $J^P={\frac{1}{2}}^+$ and $B({\frac{3}{2}}^+)$ for ground state baryons of $J^P={\frac{3}{2}}^+$. We do not mix these channels. The interaction in each block is constructed from the exchange of vector mesons obtained from the extrapolation of the local hidden gauge approach \cite{Bando:1984ej,bando1988nonlinear,Harada:2003jx,Meissner:1987ge,Nagahiro:2008cv} to the charm or bottom sector \cite{Debastiani:2017ewu,Dias:2018qhp,Dias:2019klk,Yu:2018yxl,Yu:2019yfr,Liang:2017ejq,Dong:2021bvy,Dong:2021juy,Kong:2021ohg}. The mixing of the blocks requires pion exchange, or of some other pseudoscalar, but these terms are not competitive with the vector exchange terms in the determination of the masses of the obtained states. They can contribute to the widths of the states, but in cases of many coupled channels where decay to states of lower mass can proceed via vector exchange, they are again not competitive (see appendix of Ref~\cite{Dias:2021upl}). Then the coupled channels that we consider are given below and the interaction will be considered in s-wave, which determines the $J^P$ character of the states.

\begin{table}[H]
\centering
\caption{Threshold masses (in MeV) of different channels  for $\Omega_{cc}$.}
\label{Omegacc_1}
\setlength{\tabcolsep}{8pt}
\begin{tabular}{l|cccc}
\hline \hline
  \multirow{2}*{$PB({\frac{1}{2}}^+), J^P={\frac{1}{2}}^-$} 
                        & $\Xi_{cc}\bar{K}$ 	& $\Omega_{cc}\eta$ 	& $\Xi_c D$ 	& $\Xi'_c D$         \\
            ~          & $4115$ 			& $4263$ 			& $4338$ 	& $4448$       	 \\
\hline
  \multirow{2}*{$PB({\frac{3}{2}}^+), J^P={\frac{3}{2}}^-$} 
                        & $\Xi^{*}_{cc}\bar{K}$ & $\Omega^{*}_{cc}\eta$ & $\Xi^{*}_c D$	\\
            ~          & $4168$		        & $4320$ 			& $4516$ 		  \\
\hline
  \multirow{2}*{$VB({\frac{1}{2}}^+)$, $J^P={\frac{1}{2}}^-,{\frac{3}{2}}^-$}
                        & $\Xi_{cc}\bar{K}^{*}$ 	& $\Omega_{cc}\omega$ & $\Xi_c D^{*}$ & $\Xi'_c D^{*}$    \\
            ~          & $4512$ 			& $4495$ 			& $4478$		 & $4588$	 	  \\
\hline
  \multirow{2}*{$VB({\frac{3}{2}}^+)$, $J^P={\frac{1}{2}}^-,{\frac{3}{2}}^-,{\frac{5}{2}}^-$}
                        & $\Xi^{*}_{cc}\bar{K}^{*}$ & $\Omega^{*}_{cc}\omega$ & $\Xi^{*}_c D^{*}$    \\
            ~          & $4565$ 			 & $4552$ 				& $4656$ 	 		  \\     
\hline
\hline
\end{tabular}
\end{table}

We obtain spin degenerate states in some cases. The degeneracy is expected to be broken with the consideration of pseudoscar exchange terms, but with the former considerations we expect this breaking to be small. One idea of effects expected can be seen in the splitting of the two pentaquarks states $P_{c2}$ at $4440.3$~MeV and $P_{c3}$ at $4457.3$~MeV \cite{LHCb:2019kea} from the value of the previous experiment \cite{LHCb:2015yax} combining the two states with a peak at $4450$~MeV. These states are degenerate in $J^P={\frac{1}{2}}^-,{\frac{3}{2}}^-$ for the predictions done in \cite{Wu:2010jy} mostly as a $\bar{D}^*\Sigma_c$ state, and nearly degenerate in \cite{Xiao:2013yca} where a small admixture of channels is allowed. The splitting of the states is better obtained in works considering pion exchange terms explicitly \cite{Yan:2018zdt,Yalikun:2021bfm,Wang:2022oof}.
\begin{table}[H]
\centering
\caption{Threshold masses (in MeV) of different channels  for $\Omega_{bb}$.}
\label{Omegabb_1}
\setlength{\tabcolsep}{8pt}
\begin{tabular}{l|cccc}
\hline \hline
  \multirow{2}*{$PB({\frac{1}{2}}^+), J^P={\frac{1}{2}}^-$} 
                        & $\Xi_{bb}\bar{K}$ & $\Omega_{bb}\eta$ & $\Xi_b \bar{B}$ & $\Xi'_b \bar{B}$      \\
            ~          & $10833$ 	& $10778$ 	& $11076$ 	& $11214$       	 \\
\hline
  \multirow{2}*{$PB({\frac{3}{2}}^+), J^P={\frac{3}{2}}^-$} 
                        & $\Xi^{*}_{bb}\bar{K}$ & $\Omega^{*}_{bb}\eta$ & $\Xi^{*}_b \bar{B}$\\
            ~          & $10863$ 	& $10806$ 	& $11231$		  \\
\hline
  \multirow{2}*{$VB({\frac{1}{2}}^+)$, $J^P={\frac{1}{2}}^-,{\frac{3}{2}}^-$}
                        &$\Xi_{bb}\bar{K}^{*}$ & $\Omega_{bb}\omega$ & $\Xi_b \bar{B}^{*}$ & $\Xi'_b \bar{B}^{*}$   \\
            ~          & $11230$ 	& $11010$ 	& $11122$ 	& $11260$ 	 	  \\
\hline
  \multirow{2}*{$VB({\frac{3}{2}}^+)$, $J^P={\frac{1}{2}}^-,{\frac{3}{2}}^-,{\frac{5}{2}}^-$}
                        & $\Xi^{*}_{bb}\bar{K}^{*}$ & $\Omega^{*}_{bb}\omega$ & $\Xi^{*}_b \bar{B}^{*}$   \\
            ~          & $11260$ 	& $11038$ 	& $11277$ 	 		  \\     
\hline
\hline
\end{tabular}
\end{table}

\begin{table}[H]
\centering
\caption{Threshold masses (in MeV) of different channels  for $\Omega_{bc}$.}
\label{Omegabc_1}
\setlength{\tabcolsep}{8pt}
\begin{tabular}{l|cccc}
\hline \hline
  \multirow{2}*{$PB({\frac{1}{2}}^+), J^P={\frac{1}{2}}^-$} 
                        & $\Xi_{bc}\bar{K}$ & $\Omega_{bc}\eta$  & $\Xi_b D$ & $\Xi_c \bar{B}$     \\
            ~          &  $7415$ 	& $7559$ 	& $7667$ 	& $ 7747$       	 \\
\hline
  \multirow{2}*{$PB({\frac{1}{2}}^+), J^P={\frac{1}{2}}^-$} 
                        & $\Xi'_{bc}\bar{K}$ & $\Omega'_{bc}\eta$  & $\Xi'_b D$ & $\Xi'_c \bar{B}$     \\
            ~          &  $7441$ 	& $7595$ 	& $7805$ 	& $7857$   	 \\
\hline
  \multirow{2}*{$PB({\frac{3}{2}}^+), J^P={\frac{3}{2}}^-$}
                        &$\Xi^{*}_{bc}\bar{K}$ & $\Omega^{*}_{bc}\eta$  & $\Xi^{*}_b D$ & $\Xi^{*}_c \bar{B}$   \\
            ~          &  $7466$ 	& $7614$ 	& $7822$ 	& $7925$	 	  \\
\hline
  \multirow{2}*{$VB({\frac{1}{2}}^+)$, $J^P={\frac{1}{2}}^-,{\frac{3}{2}}^-$}
                        &$\Xi_{bc}\bar{K}^{*}$ & $\Omega_{bc}\omega$  & $\Xi_b D^{*}$ & $\Xi_c \bar{B}^{*}$  \\
            ~          &  $7812$ 	& $7791$ 	& $7807$ 	& $7793$	 	  \\
\hline
  \multirow{2}*{$VB({\frac{1}{2}}^+)$, $J^P={\frac{1}{2}}^-,{\frac{3}{2}}^-$}
                        &$\Xi'_{bc}\bar{K}^{*}$ & $\Omega'_{bc}\omega$  & $\Xi'_b D^{*}$ & $\Xi'_c \bar{B}^{*}$  \\
            ~          &  $7838$ 	& $7827$ 	& $7945$ 	& $7903$	      \\
\hline
  \multirow{2}*{$VB({\frac{3}{2}}^+)$, $J^P={\frac{1}{2}}^-,{\frac{3}{2}}^-,{\frac{5}{2}}^-$}
                        &$\Xi^{*}_{bc}\bar{K}^{*}$ & $\Omega^{*}_{bc}\omega$ & $\Xi^{*}_b D^{*}$& $\Xi^{*}_c\bar{B}^{*}$  \\
            ~          & $7863$ 	& $7846$ 	& $7962$ 	& $7971$	     \\           
\hline
\hline
\end{tabular}
\end{table}

\subsection{Baryon wave functions}  

The flavor of the pseudoscalar or vector mesons is trivial as they are $q_i\bar{q}_j$ states, with $q_i=u,d,s,c,b$ quarks. The baryon states require more care. We follow the procedure of Ref.~\cite{Roberts:2007ni} and single out the heaviest quark, then the symmetry or antisymmetry is imposed on the two lighter quarks, and the spin wave function is then chosen accordingly to have the wave function symmetric in spin-flavor for this couple of quarks, the color implementing the antisymmetry of the wave function.
In Table~\ref{Baryon_wf} we show explicitly the wave functions of flavor and spin taken for all the baryon states needed in our work, and where the spin wave functions within are defined, for the particular case $S_z=+1/2$, as:

\begin{eqnarray}
 \chi_{MS}(12)&=& \frac{1}{\sqrt{6}}(\uparrow\downarrow\uparrow+\downarrow\uparrow\uparrow-2\uparrow\uparrow\downarrow)\\ \nonumber
  \chi_{MS}(23)&=& \frac{1}{\sqrt{6}}(\uparrow\downarrow\uparrow+\uparrow\uparrow\downarrow-2\downarrow\uparrow\uparrow)\\ \nonumber
    \chi_{MA}(23)&=& \frac{1}{\sqrt{2}}(\uparrow\uparrow\downarrow-\uparrow\downarrow\uparrow). \\ \nonumber
\end{eqnarray}

\begin{table}[H]
\centering
 \caption{
 Wave functions of baryon states.}
 \label{Baryon_wf}
\setlength{\tabcolsep}{40pt}
\begin{tabular}{cccc}
\hline
\hline

State   & $I,J$ & flavor & spin\\
\hline
 $\Xi^{++}_{cc}$  & $1/2,1/2$ & $ccu$ & $\chi_{MS}(12)$ \\
 $\Xi^{+}_{cc}$  & $1/2,1/2$ & $ccd$ & $\chi_{MS}(12)$\\
$\Omega^{+}_{cc}$& $0,1/2$ & $ccs$ & $\chi_{MS}(12)$\\
  $\Xi^{+}_c$    & $1/2,1/2$ & $\frac{1}{\sqrt{2}}c(us-su)$ & $\chi_{MA}(23)$\\
$\Xi^{0}_c$  & $1/2,1/2$ & $\frac{1}{\sqrt{2}}c(ds-sd)$ & $\chi_{MA}(23)$\\
$\Xi^{'+}_c$  & $1/2,1/2$ & $\frac{1}{\sqrt{2}}c(us+su)$ & $\chi_{MS}(23)$\\
$\Xi^{'0}_c$ & $1/2,1/2$ & $\frac{1}{\sqrt{2}}c(ds+sd)$ & $\chi_{MS}(23)$ \\
$\Omega^{0}_{c}$ & $0,1/2$ & $css$ & $\chi_{MS}(23)$ \\
     \hline
 $\Xi^{0}_{bb}$  & $1/2,1/2$ & $bbu$ & $\chi_{MS}(12)$ \\
 $\Xi^{-}_{bb}$  & $1/2,1/2$ & $bbd$ & $\chi_{MS}(12)$\\
$\Omega^{-}_{bb}$& $0,1/2$ & $bbs$ & $\chi_{MS}(12)$\\
  $\Xi^{0}_b$    & $1/2,1/2$ & $\frac{1}{\sqrt{2}}b(us-su)$ & $\chi_{MA}(23)$\\
$\Xi^{-}_b$  & $1/2,1/2$ & $\frac{1}{\sqrt{2}}b(ds-sd)$ & $\chi_{MA}(23)$\\
$\Xi^{'0}_b$  & $1/2,1/2$ & $\frac{1}{\sqrt{2}}b(us+su)$ & $\chi_{MS}(23)$\\
$\Xi^{'-}_b$ & $1/2,1/2$ & $\frac{1}{\sqrt{2}}b(ds+sd)$ & $\chi_{MS}(23)$ \\
$\Omega^{-}_{b}$ & $0,1/2$ & $bss$ & $\chi_{MS}(23)$ \\
 \hline
$\Omega^{0}_{bc}$& $0,1/2$ & $\frac{1}{\sqrt{2}}b(cs-sc)$ & $\chi_{MA}(23)$\\
$\Omega^{'0}_{bc}$& $0,1/2$ & $\frac{1}{\sqrt{2}}b(cs+sc)$ & $\chi_{MS}(23)$\\
$\Xi^{+}_{bc}$    & $1/2,1/2$ & $\frac{1}{\sqrt{2}}b(cu-uc)$ & $\chi_{MA}(23)$\\
$\Xi^{'+}_{bc}$    & $1/2,1/2$ & $\frac{1}{\sqrt{2}}b(cu+uc)$ & $\chi_{MS}(23)$\\
$\Xi^{0}_{bc}$    & $1/2,1/2$ & $\frac{1}{\sqrt{2}}b(cd-dc)$ & $\chi_{MA}(23)$\\
$\Xi^{'0}_{bc}$    & $1/2,1/2$ & $\frac{1}{\sqrt{2}}b(cd+dc)$ & $\chi_{MS}(23)$\\
\hline
 $\Xi^{*++}_{cc}$  & $1/2,3/2$ & $ccu$ & $\chi_{S}$ \\
 $\Xi^{*+}_{cc}$  & $1/2,3/2$ & $ccd$ & $\chi_{S}$\\
$\Omega^{*+}_{cc}$& $0,3/2$ & $ccs$ & $\chi_{S}$\\
$\Xi^{*+}_c$    & $1/2,3/2$ & $\frac{1}{\sqrt{2}}c(us+su)$ & $\chi_{S}$\\
$\Xi^{*0}_c$  & $1/2,3/2$ & $\frac{1}{\sqrt{2}}c(ds+sd)$ & $\chi_{S}$\\
$\Omega^{*0}_{c}$ & $0,3/2$ & $css$ & $\chi_{S}$ \\
\hline
$\Xi^{*0}_{bb}$  & $1/2,3/2$ & $bbu$ & $\chi_{S}$ \\
 $\Xi^{*-}_{bb}$ & $1/2,3/2$ & $bbd$ & $\chi_{S}$\\
$\Omega^{*-}_{bb}$& $0,3/2$ & $bbs$ & $\chi_{S}$\\
$\Xi^{*0}_b$    & $1/2,3/2$ & $\frac{1}{\sqrt{2}}b(us+su)$ & $\chi_{S}$\\
$\Xi^{*-}_b$  & $1/2,3/2$ & $\frac{1}{\sqrt{2}}b(ds+sd)$ & $\chi_{S}$\\
$\Omega^{*-}_{b}$ & $0,3/2$ & $bss$ & $\chi_{S}$ \\
\hline
$\Xi^{*+}_{bc}$    & $1/2,3/2$ & $\frac{1}{\sqrt{2}}b(cu+uc)$ & $\chi_{S}$\\
$\Xi^{*0}_{bc}$    & $1/2,3/2$ & $\frac{1}{\sqrt{2}}b(cd+dc)$ & $\chi_{S}$\\
$\Omega^{*0}_{bc}$& $0,3/2$ & $\frac{1}{\sqrt{2}}b(cs+sc)$ & $\chi_{S}$\\
\hline
\hline
\end{tabular}
\end{table}

We also must consider the isospin combinations of the states. For this we need to express our phase convention 
$\begin{Bmatrix} \bar{K}^0 \\ -K^- \end{Bmatrix}, \begin{Bmatrix} D^+\\ -D^0  \end{Bmatrix}, \begin{Bmatrix} \bar{B}^0 \\ -B^-   \end{Bmatrix}, \begin{Bmatrix} \Xi_{c}^+\\ \Xi_{c}^0 \end{Bmatrix}, \begin{Bmatrix} \Xi_{b}^0\\ \Xi_{b}^- \end{Bmatrix}, \begin{Bmatrix} \Xi_{cc}^{++}\\ \Xi_{cc}^+  \end{Bmatrix}, \begin{Bmatrix} \Xi_{bb}^{0}\\ \Xi_{bb}^-  \end{Bmatrix}, \begin{Bmatrix}\Xi_{bc}^{+}\\ \Xi_{bc}^0   \end{Bmatrix}$, and then the isospin wave functions are given by:

\begin{eqnarray}
\label{ISO_WF}
| \Xi^{(*)}_{cc} \bar{K}^{(*)}, I=0\rangle &=& -\frac{1}{\sqrt{2}} ( |\Xi_{cc}^{(*)++} K^{(*)-}\rangle + |\Xi_{cc}^{(*)+} \bar{K}^{(*)0}\rangle )\\ \nonumber
| \Omega^{(*)}_{cc} \eta , I=0\rangle &=& |\Omega^{(*)+}_{cc} \eta \rangle \\ \nonumber
| \Omega^{(*)}_{cc} \omega , I=0\rangle &=& |\Omega^{(*)+}_{cc} \omega \rangle \\ \nonumber
| \Xi^{(*)}_{c}D^{(*)}, I=0\rangle &=&-\frac{1}{\sqrt{2}} ( |\Xi_{c}^{(*)+} D^{(*)0}\rangle + |\Xi_{c}^{(*)0} D^{(*)+}\rangle ) \\ \nonumber
| \Xi^{'}_{c}D^{(*)}, I=0\rangle &=&-\frac{1}{\sqrt{2}} ( |\Xi_{c}^{'+} D^{(*)0}\rangle + |\Xi_{c}^{'0} D^{(*)+}\rangle ) \\ \nonumber
| \Xi^{(*)}_{bb} \bar{K}^{(*)}, I=0\rangle &=& -\frac{1}{\sqrt{2}} ( |\Xi_{bb}^{(*)0} K^{(*)-}\rangle + |\Xi_{bb}^{(*)-} \bar{K}^{(*)0}\rangle )\\ \nonumber
| \Omega^{(*)}_{bb} \eta , I=0\rangle &=& |\Omega^{(*)-}_{bb} \eta \rangle \\ \nonumber
| \Omega^{(*)}_{bb} \omega , I=0\rangle &=& |\Omega^{(*)-}_{bb} \omega \rangle \\ \nonumber
| \Xi_{b}^{(*)} \bar{B}^{(*)}, I=0\rangle &=& -\frac{1}{\sqrt{2}} ( |\Xi_{b}^{(*)0} B^{(*)-}\rangle + |\Xi_{b}^{(*)-} \bar{B}^{(*)0}\rangle )\\ \nonumber
| \Xi_{b}^{'} \bar{B}^{(*)}, I=0\rangle &=& -\frac{1}{\sqrt{2}} ( |\Xi_{b}^{'0} B^{(*)-}\rangle + |\Xi_{b}^{'-} \bar{B}^{(*)0}\rangle )\\ \nonumber
| \Xi_{bc}^{(*)} \bar{K}^{(*)}, I=0\rangle &=& -\frac{1}{\sqrt{2}} ( |\Xi_{bc}^{(*)+} K^{(*)-}\rangle + |\Xi_{bc}^{(*)0} \bar{K}^{(*)0}\rangle )\\ \nonumber
| \Xi_{bc}^{'} \bar{K}^{(*)}, I=0\rangle &=& -\frac{1}{\sqrt{2}} ( |\Xi_{bc}^{'+} K^{(*)-}\rangle + |\Xi_{bc}^{'0} \bar{K}^{(*)0}\rangle )\\ \nonumber
| \Omega^{(*)}_{bc} \eta , I=0\rangle &=& |\Omega^{(*)0}_{bc} \eta \rangle \\ \nonumber
| \Omega^{'}_{bc} \eta , I=0\rangle &=& |\Omega^{'0}_{bc} \eta \rangle \\ \nonumber
| \Omega^{(*)}_{bc} \omega , I=0\rangle &=& |\Omega^{(*)0}_{bc} \omega \rangle \\ \nonumber
| \Omega^{'}_{bc} \omega , I=0\rangle &=& |\Omega^{'0}_{bc} \omega \rangle \\ \nonumber
| \Xi_{b}^{(*)}D^{(*)}, I=0\rangle &=&-\frac{1}{\sqrt{2}} ( |\Xi_{b}^{(*)0} D^{(*)0}\rangle + |\Xi_{b}^{(*)-} D^{(*)+}\rangle ) \\ \nonumber
| \Xi_{b}^{'}D^{(*)}, I=0\rangle &=&-\frac{1}{\sqrt{2}} ( |\Xi_{b}^{'0} D^{(*)0}\rangle + |\Xi_{b}^{'-} D^{(*)+}\rangle ) \\ \nonumber
| \Xi_{c}^{(*)} \bar{B}^{(*)}, I=0\rangle &=& -\frac{1}{\sqrt{2}} ( |\Xi_{c}^{(*)+} B^{(*)-}\rangle + |\Xi_{c}^{(*)0} \bar{B}^{(*)0}\rangle )\\ \nonumber
|\Xi_{c}^{'} \bar{B}^{(*)}, I=0\rangle &=& -\frac{1}{\sqrt{2}} ( |\Xi_{c}^{'+} B^{(*)-}\rangle + |\Xi_{c}^{'0} \bar{B}^{(*)0}\rangle ).\\ \nonumber
\end{eqnarray}
    
\subsection{Interaction between coupled channels}

\begin{figure}[H]   
  \centering
  \includegraphics[width=4.5cm]{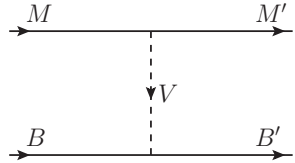}
  \caption{Diagrammatic representation of the interaction $MB\to M^\prime B^\prime$ through the exchange of vector mesons.
                The $M (M^\prime)$ and $B (B^\prime)$ are the initial (final) meson and baryon states, respectively, while $V$ stands for 
                the vector meson exchanged.
              }
  \label{fig1}
\end{figure}

As mentioned above, we use vector exchange from the extension of the local hidden gauge approach between mesons 
and baryons as shown in the Fig.~\ref{fig1}. 
The $VMM^\prime$ vertex has two types for our set of states, $VPP$ ($V\equiv$ vector, $P\equiv$ pseudoscalar) and 
$VVV$, which are described by the following Lagrangians 
\begin{align}
    \mathcal{L}_{\mathrm{VPP}} &= -i g\left\langle\left[P, \partial_{\mu} P\right] V^{\mu}\right\rangle, \label{eq-1}\\
    \mathcal{L}_{\mathrm{VVV}} &=  i g\left\langle\left(V^{\mu} \partial_{\nu} V_{\mu}-\partial_{\nu} V^{\mu}
                                     V_{\mu} \right) V^{\nu}\right\rangle. \label{eq-2}    
\end{align}
The coupling  $g=\frac{m_V}{2f_\pi}$ with $m_V=800$ MeV and the pion decay constant $f_\pi=93$ MeV. And the $P$ or $V$ 
above are the $q_i\bar{q}_j$ matrices written in terms of mesons and the symbol $\langle\cdot \cdot \cdot\rangle$ means the 
trace for the matrices. We must recall that, while $q_i\bar{q}_j$ are SU$(4)$ or SU$(5)$ matrices, the vertices of Eqs.~(\ref{eq-1}) and (\ref{eq-2}) 
only use the overlap of $q\bar{q}$ in the external mesons and the exchanged vectors, hence the use of the SU$(4)$, SU$(5)$ symmetry 
is superfluous~\cite{Sakai:2017avl}.

The matrices $P$ and $V$ that we need are given by 
\begin{equation}
 \label{eq:matP_charm} 
  P = \begin{pmatrix}
          \frac{1}{\sqrt{2}}\pi^0 + \frac{1}{\sqrt{3}} \eta + \frac{1}{\sqrt{6}}\eta' & \pi^+ & K^+ & \bar{D}^0 \\
          \pi^- & -\frac{1}{\sqrt{2}}\pi^0 + \frac{1}{\sqrt{3}} \eta + \frac{1}{\sqrt{6}}\eta' & K^0 & D^- \\
          K^- & \bar{K}^0 & -\frac{1}{\sqrt{3}} \eta + \sqrt{\frac{2}{3}}\eta' & D_s^- \\
          D^0  & D^+ & D_s^+ & \eta_c
       \end{pmatrix},
\end{equation}
\begin{equation}
  \label{eq:matV_charm}
   V = \begin{pmatrix}
             \frac{1}{\sqrt{2}}\rho^0 + \frac{1}{\sqrt{2}} \omega & \rho^+ & K^{* +} & \bar{D}^{* 0} \\
             \rho^- & -\frac{1}{\sqrt{2}}\rho^0 + \frac{1}{\sqrt{2}} \omega  & K^{* 0} & \bar{D}^{* -} \\
             K^{* -} & \bar{K}^{* 0}  & \phi & D_s^{* -} \\
             D^{* 0} & D^{* +} & D_s^{* +} & J/\psi
        \end{pmatrix},
\end{equation}
for the mesons in the charm sector, and
\begin{equation}
 \label{eq:matP_bottom}
    P = \begin{pmatrix}
            \frac{1}{\sqrt{2}}\pi^0 + \frac{1}{\sqrt{3}} \eta + \frac{1}{\sqrt{6}}\eta' & \pi^+ & K^+ & B^+ \\
            \pi^- & -\frac{1}{\sqrt{2}}\pi^0 + \frac{1}{\sqrt{3}} \eta + \frac{1}{\sqrt{6}}\eta' & K^0 & B^0 \\
            K^- & \bar{K}^0 & -\frac{1}{\sqrt{3}} \eta + \sqrt{\frac{2}{3}}\eta' & B_s^0 \\
            B^-  & \bar B^0 & \bar B_s^0 & \eta_b
         \end{pmatrix},
\end{equation}
\begin{equation}
\label{eq:matV_bottom}
   V = \begin{pmatrix}
            \frac{1}{\sqrt{2}}\rho^0 + \frac{1}{\sqrt{2}} \omega & \rho^+ & K^{* +} & B^{* +} \\
            \rho^- & -\frac{1}{\sqrt{2}}\rho^0 + \frac{1}{\sqrt{2}} \omega  & K^{* 0} & B^{* 0} \\
           K^{* -} & \bar{K}^{* 0}  & \phi & B_s^{* 0} \\
           B^{* -} & \bar B^{* 0} & \bar B_s^{* 0} & \Upsilon
         \end{pmatrix},
\end{equation}
for the bottom sector.

We also note that we evaluate the $VVV$ interaction assuming that the external three momenta are small versus the vector mass and 
then can be neglected. In this case the vector $V^\nu$ in Eq.~(\ref{eq-2}) cannot be the external vector 
since $\epsilon^0=0$ for the vector at rest, the $\nu$ are spatial components and $\partial_\nu$ will give three vectors which 
are null. Then $V^\nu$ corresponds to the exchanged vector. Eq.~(\ref{eq-2}) is then
equivalent to Eq.~(\ref{eq-1}) substituting the $P$ of a given $q\bar{q}$ by the corresponding 
$V$ and adding the $-\epsilon^\mu\epsilon^\prime_\mu=\vec \epsilon \cdot \vec \epsilon\,^\prime$ factor for the polarization of the 
two external vectors.

The lower vertex of Fig.~\ref{fig1} is rendered easy to evaluate using the wave functions of Table~\ref{Baryon_wf}. 
Rather than using effective Lagrangians 
in terms of the mesons and baryons which require extension of chiral Lagrangians from SU$(3)$ to SU$(4)$ or 
SU$(5)$~\cite{Hofmann:2005sw,Mizutani:2006vq,Montana:2017kjw}, we write the operator in terms of quarks 
and sandwich it with the baryon wave functions of Table~\ref{Baryon_wf}~\cite{Debastiani:2017ewu}. The vertex is given by 
\begin{eqnarray}
  \widetilde{\mathcal{L}}_{\mathrm{VBB}} \equiv g q \bar{q}(V),
  \label{eq:Lag_VBB}
\end{eqnarray}
where $q\bar{q}(V)$ is the vector wave function in terms of quarks, hence 
\begin{equation}
   \widetilde{{\cal L}}_{VBB} \equiv g \left\{ 
         \begin{aligned}
           & \frac{1}{\sqrt{2}} (u\bar{u} - d\bar{d}),\quad \rho^0 \\
           & \frac{1}{\sqrt{2}} (u\bar{u} + d\bar{d}),\quad \omega \\
           & \qquad~~ s\bar{s}, \qquad\quad~\, \phi \\
         \end{aligned} \right\},
     \label{eq:L_VBB}
\end{equation}
in the neutral light vector exchange. It is worth mentioning that the exchange of light vector mesons obtained using SU$(4)$ symmetry 
in \cite{Hofmann:2005sw,Mizutani:2006vq,Montana:2017kjw} coincides with the formalism of~\cite{Debastiani:2017ewu} since the heavy quarks are spectators and only the light quarks play a role, hence one is not making use of SU$(4)$ symmetry, but only of its 
SU$(3)$ subgroup. Since the exchange of light vector mesons provide the dominant contribution, it is not surprising to see that the results of~\cite{Montana:2017kjw} and~\cite{Debastiani:2017ewu} are very similar. The other point worth mentioning is that for these 
dominant terms, since the heavy quarks are spectators, the interaction does not depend upon them and then the heavy quark 
symmetries~\cite{Manohar:1983md} are automatically fulfilled.

An example of how the $MB\to M^\prime B^\prime$ potential is obtained combining the two vertices of Fig.~\ref{fig1} and the vector 
propagator is shown in the Appendix of Ref.~\cite{Debastiani:2017ewu}. It is very easy and we refrain from repeating it here.

The evaluation of the transition potential $V_{ij}$ from one channel to another in the blocks we have selected is then straightforward,
but we can simplify the calculation with the following observation~\cite{Dias:2019klk}. Since the $VBB$ vertex is spin independent 
we classify the state in blocks that have $\chi_{MS}(23)$, $\chi_{MA}(23)$ and $\chi_{S}$. Since $\chi_{MS}(12)$ have overlap with 
$\chi_{MS}(23)$ and $\chi_{MA}(23)$, we must keep the states of $\chi_{MS}(12)$ in the blocks of $\chi_{MS}(23)$ and 
$\chi_{MA}(23)$. The overlap of these spin functions are given by 
\begin{align}
  \langle \chi_{MS}(12)| \chi_{MS}(23) \rangle &= -\frac{1}{2}, \\
  \langle \chi_{MS}(12)| \chi_{MA}(23) \rangle & =-\frac{\sqrt{3}}{2}.
\end{align}
However, even if the $\Xi_c^\prime D$ and $\Xi_c D$ states do not mix in the calculation of the potential, in the $T$ matrix they will 
mix through the intermediate $\Xi_{cc} \bar{K}$ and $\Omega_{cc}\eta$ states. Hence, we write the matrix $V_{ij}$ for all the 
states of $PB$. This said, we have the following blocks.

\subsection{Spin blocks for $\Omega_{cc}$ state}   

\begin{itemize}
  \item [i)] $PB (\frac12^+)$ channels:
                $\Xi_{cc}\bar{K}$, $\Omega_{cc}\eta$, $\Xi_cD$, and $\Xi_c^\prime D$.

  \item [ii)] $PB (\frac32^+)$ channels:
                $\Xi_{cc}^\ast\bar{K}$, $\Omega_{cc}^\ast\eta$, $\Xi_c^\ast D$
 
  \item [iii)] $VB (\frac12^+)$ channels:
                 $\Xi_{cc}\bar{K}^\ast$, $\Xi_cD^\ast$, $\Omega_{cc}\omega$, and $\Xi_c^\prime D^\ast$.
  
  \item [iv)] $VB (\frac32^+)$ channels:
                 $\Xi_{cc}^\ast\bar{K}^\ast$, $\Omega_{cc}^\ast\omega$, $\Xi_c^\ast D^\ast$.

\end{itemize}

The interaction obtained for the mechanism of Fig.~\ref{fig1}  is always of the type
\begin{eqnarray}
  \label{eq:def_Vij}
     V_{ij}= -\frac{1}{4f_\pi^2}(p_1^0+p^{0}_3)C_{ij} \, ,
\end{eqnarray}
where $p_1^0$, $p^{0}_3$ are the energies of the initial and final mesons, respectively. The coefficient $C_{ij}$ are then evaluated 
and we find the following Tables~\ref{coeff_cc_PB}-\ref{coeff_cc_VB1}
 for the blocks described before. The $\lambda$ below is a suppression factor of the order 
$m^2_V/m^2_{D^*}$ coming from the exchange of a $D^\ast$ rather than a light vector. Following Ref.~\cite{Debastiani:2017ewu} 
we take the value $\lambda=0.25$ in the numerical calculations.

\begin{table}[H]
\centering
 \caption{Coefficients $C_{ij}$ for the PB sector with $J^P = \frac{1}{2}^{-}$.}
 \label{coeff_cc_PB}
\setlength{\tabcolsep}{6.5pt}
\begin{tabular}{l|cccc}
\hline
\hline  
          ~            &$\Xi_{cc}\bar{K}\;$    & $\Omega_{cc}\eta$       & $\Xi_c D$                               & $\Xi'_c D$ \\
\hline
  $\Xi_{cc}\bar{K}$       & $2$   & $\frac{2\sqrt{2}}{\sqrt{3}}$   & $\frac{-\sqrt{3}}{2\sqrt{2}}\lambda$  & $\frac{1}{2\sqrt{2}}\lambda$   \\
  $\Omega_{cc}\eta$   &          & $0$                                       & $-\frac{1}{2}\lambda$                        & $\frac{-1}{2\sqrt{3}}\lambda$  \\
  $\Xi_c D$                   &          &                                              & $2$                                                    & 0                    \\
  $\Xi'_c D$                  &          &                                              &                                                            & $2$           \\
\hline
\hline
\end{tabular}
\end{table}

\begin{table}[H]
\centering
 \caption{Coefficients $C_{ij}$ for the VB sector with $J^P = \frac{1}{2}^{-}, \frac{3}{2}^{-}$.}
 \label{coeff_cc_VB}
\setlength{\tabcolsep}{6.5pt}
\begin{tabular}{l|cccc}
\hline
\hline  
  ~  				 & $\Xi_c D^{*}$        & $\Omega_{cc}\omega$ & $\Xi_{cc}\bar{K}^{*}$    & $\Xi'_c D^{*}$ \\
\hline
   $\Xi_c D^{*}$                &  $2$               & $\frac{-\sqrt{3}}{2\sqrt{2}}\lambda$ &   $\frac{-\sqrt{3}}{2\sqrt{2}}\lambda$ & 0   \\
   $\Omega_{cc}\omega$ &                      &$0$             & $1$                             & $\frac{-1}{2\sqrt{2}}\lambda$ 			\\
   $\Xi_{cc}\bar{K}^{*}$     &                       &                  & $2$                             & $\frac{1}{2\sqrt{2}}\lambda$ 			\\
   $\Xi'_c D^{*}$                &                       &                    &           			     & $2$  							 \\
\hline
\hline
\end{tabular}
\end{table}

\begin{table}[H]
\centering
 \caption{Coefficients $C_{ij}$ for the PB sector with $J^P = \frac{3}{2}^{-}$.}
 \label{coeff_cc_PB1}
\setlength{\tabcolsep}{6.5pt}
\begin{tabular}{l|ccc}
\hline
\hline
      ~                             & $\Xi^{*}_{cc}\bar{K}$  & $\Omega^{*}_{cc}\eta$     & $\Xi^{*}_c D$                       \\
\hline
  $\Xi^{*}_{cc}\bar{K}$    &      		 $2$             & $\frac{2\sqrt{2}}{\sqrt{3}}$ & $\frac{1}{\sqrt{2}}\lambda$    \\
  $\Omega^{*}_{cc}\eta$ &  				    & $0$ 					  & $\frac{1}{\sqrt{3}}\lambda$    \\
  $\Xi^{*}_c D$ 		& 			            & 						  & $2$                                          \\ 
\hline
\hline
\end{tabular}
\end{table}

\begin{table}[H]
\centering
 \caption{Coefficients $C_{ij}$ for the VB sector with $J^P = \frac{1}{2}^{-}, \frac{3}{2}^{-},  \frac{5}{2}^{-}$.}
 \label{coeff_cc_VB1}
\setlength{\tabcolsep}{6.5pt}
\begin{tabular}{l|ccc}
\hline
\hline
  ~ 						& $\Omega^{*}_{cc}\omega$  & $\Xi^{*}_{cc}\bar{K}^{*}$ & $\Xi^{*}_c D^{*}$     \\
\hline
   $\Omega^{*}_{cc}\omega$  &$0$ 					& $1$ 			& $\frac{1}{\sqrt{2}}\lambda$   \\
   $\Xi^{*}_{cc}\bar{K}^{*}$      &  						& $2$ 			& $\frac{1}{\sqrt{2}}\lambda$   \\
    $\Xi^{*}_c D^{*}$                 &  						&  				& $2$  					 \\ 
\hline
\hline
\end{tabular}
\end{table}

\subsection{Spin blocks for $\Omega_{bb}$ states}

In these sectors we have taken zero the terms that go with the exchange of $B^\ast$, since $m^2_V/m^2_{B^\ast}$ is negligible. 
We follow the same steps as before and find the $C_{ij}$ coefficients and the channels belonging to each block as shown in 
Tables~\ref{coeff_bb_PB}-\ref{coeff_bb_VB1}.

\begin{table}[H]
\centering
 \caption{Coefficients $C_{ij}$ for the PB sector with $J^P = \frac{1}{2}^{-}$.}
 \label{coeff_bb_PB}
\setlength{\tabcolsep}{6.5pt}
\begin{tabular}{l|cccc}
\hline
\hline
~                      & $\Omega_{bb}\eta$   & $\Xi_{bb}\bar{K}$   &  $\Xi_b \bar{B}$   & $\Xi'_b \bar{B}$  \\
\hline
  $\Omega_{bb}\eta$  & $0$     & $\frac{2\sqrt{2}}{\sqrt{3}}$  & $0$     & $0$      \\
  $\Xi_{bb}\bar{K}$      &            & $2$					 & $0$     & $0$      \\
  $\Xi_b \bar{B}$         &            &                                             & $2$     & $0$       \\
  $\Xi'_b \bar{B}$        &            &  						 &            & $2$        \\ 
\hline
\hline
\end{tabular}
\end{table}

\begin{table}[H]
\centering
 \caption{Coefficients $C_{ij}$ for the VB sector with $J^P = \frac{1}{2}^{-}, \frac{3}{2}^{-}$.}
 \label{coeff_bb_VB}
 \setlength{\tabcolsep}{6.5pt}
\begin{tabular}{c|cccccccc}
\hline
\hline
 ~  & $\Omega_{bb}\omega$  & $\Xi_b \bar{B}^{*}$ & $\Xi_{bb}\bar{K}^{*}$  & $\Xi'_b \bar{B}^{*}$ \\
\hline
  $\Omega_{bb}\omega$     &  $0$    & $0$    & $1$   & $0$    \\
  $\Xi_b \bar{B}^{*}$            &             &$2$     & $0$   & $0$    \\
  $\Xi_{bb}\bar{K}^{*}$         &            &            &$2$    & $0$    \\
  $\Xi'_b \bar{B}^{*}$           & 		   &  		 &	     & $2$     \\
\hline
\hline
\end{tabular}
\end{table}

\begin{table}[H]
\centering
 \caption{Coefficients $C_{ij}$ for the PB sector with $J^P = \frac{3}{2}^{-}$.}
 \label{coeff_bb_PB1}
 \setlength{\tabcolsep}{6.5pt}
 \begin{tabular}{l|ccc}
\hline
\hline
~                   & $\Omega^{*}_{bb}\eta$ & $\Xi^{*}_{bb}\bar{K}$  & $\Xi^{*}_b \bar{B}$   \\
\hline
 $\Omega^{*}_{bb}\eta$  &  $0$ 	& $\frac{2\sqrt{2}}{\sqrt{3}}$ & $0$ 		 \\
 $\Xi^{*}_{bb}\bar{K}$     & 		& $2$				      & $0$  		 \\
 $\Xi^{*}_b \bar{B}$        &  		&  					      &  $2$		  \\
\hline
\hline
\end{tabular}
\end{table}

\begin{table}[H]
\centering
 \caption{Coefficients $C_{ij}$ for the VB sector with $J^P = \frac{1}{2}^{-}, \frac{3}{2}^{-},  \frac{5}{2}^{-}$.}
 \label{coeff_bb_VB1}
\setlength{\tabcolsep}{6.5pt}
\begin{tabular}{l|ccc}
\hline
\hline
  ~                           & $\Omega^{*}_{bb}\omega$ &  $\Xi^{*}_{bb}\bar{K}^{*}$ & $\Xi^{*}_b \bar{B}^{*}$ \\
\hline
  $\Omega^{*}_{bb}\omega$    & $0$    &$1$     & $0$   	 \\
  $\Xi^{*}_{bb}\bar{K}^{*}$        &           & $2$    & $0$		 \\
  $\Xi^{*}_b \bar{B}^{*}$            &           &           & $2$	        	 \\
\hline
\hline
\end{tabular}
\end{table}

\subsection{Spin blocks of $\Omega_{bc}$} 

The $C_{ij}$ coefficients and the channels belonging to each block are shown in Tables~\ref{coeff_bc_PB}-\ref{coeff_bc_VB1}.

\begin{table}[H]
\centering
 \caption{Coefficients $C_{ij}$ for the PB sector with $J^P = \frac{1}{2}^{-}$.}
 \label{coeff_bc_PB}
\setlength{\tabcolsep}{6.5pt}
\begin{tabular}{l|ccccccccccc}
\hline
\hline
~ & $\Xi_{bc}\bar{K}$ & $\Xi'_{bc}\bar{K}$  & $\Omega_{bc}\eta$ & $\Omega'_{bc}\eta$ 
                & $\Xi_b D$ & $\Xi_c \bar{B}$  & $\Xi'_b D$ & $\Xi'_c \bar{B}$      \\
\hline
   $\Xi_{bc}\bar{K}$ & $2$ & $0$ & $\frac{2\sqrt{2}}{\sqrt{3}}$ & $0$  & $\lambda$ & $0$ & $0$ & $0$ \\
  $\Xi'_{bc}\bar{K}$ &        & $2$ & $0$& $\frac{2\sqrt{2}}{\sqrt{3}}$ & $0$ &$0$ & $-\lambda$ & $0$   \\
 $\Omega_{bc}\eta$ &       &        & $0$& $0$ & $\frac{\sqrt{2}}{\sqrt{3}}\lambda$ & $0$& $0$ & $0$     \\
 $\Omega'_{bc}\eta$ &      &        &      & $0$ & $0$& $0$ & $\frac{\sqrt{2}}{\sqrt{3}}\lambda$& $0$       \\
  $\Xi_b D$                &       &        &      &        & $2$& $0$ & $0$ & $0$ 						       \\
  $\Xi_c \bar{B}$        &       &        &      &        &      & $2$ & $0$ & $0$ 							\\
  $\Xi'_b D$               &        &        &      &        &      &        & $2$  & $0$							\\
  $\Xi'_c \bar{B}$      &        &        &       &        &       &       &        &  $2$							\\
\hline
\hline
\end{tabular}
\end{table}

\begin{table}[H]
\centering
 \caption{Coefficients $C_{ij}$ for the VB sector with $J^P = \frac{1}{2}^{-}, \frac{3}{2}^{-}$.}
 \label{coeff_bc_VB}
\setlength{\tabcolsep}{6.5pt}
\begin{tabular}{l|ccccccccccc}
\hline
\hline
    ~  &  $\Omega_{bc}\omega$ & $\Xi_c \bar{B}^{*}$ &  $\Xi_b D^{*}$  & $\Xi_{bc}\bar{K}^{*}$  
             & $\Omega'_{bc}\omega$  & $\Xi'_{bc}\bar{K}^{*}$  & $\Xi'_c \bar{B}^{*}$  & $\Xi'_b D^{*}$       \\
\hline
  $\Omega_{bc}\omega$  & $0$  & $0$  & $\lambda$& $1$& $0$ & $0$ &$0$& $0$           \\
  $\Xi_c \bar{B}^{*}$         &          &$2$  & $0$& $0$& $0$& $0$ & $0$ &$0$                        \\
  $\Xi_b D^{*}$                  &          &        & $2$& $\lambda$& $0$& $0$ & $0$ &$0$  		 \\
  $\Xi_{bc}\bar{K}^{*}$       &         & 	& 	  & $2$  & $0$ & $0$ &$0$& $0$             	 \\
  $\Omega'_{bc}\omega$  &         &        &      &        & $0$ & $1$ &$0$& $\lambda$  		 \\
  $\Xi'_{bc}\bar{K}^{*}$      &  	     &		&      &        &        &  $2$ &$0$& $-\lambda$		 \\
  $\Xi'_c \bar{B}^{*}$          &        &		&  	&	      &       &         &  $2$  &$0$         	  \\
  $\Xi'_b D^{*}$                  &         &        &      &         &       &         &           & $2$      	          \\
\hline
\hline
\end{tabular}
\end{table}

\begin{table}[H]
\centering
 \caption{Coefficients $C_{ij}$ for the PB sector with $J^P = \frac{3}{2}^{-}$.}
 \label{coeff_bc_PB1}
\setlength{\tabcolsep}{6.5pt}
\begin{tabular}{l|cccccccc}
\hline
\hline
~    & $\Xi^{*}_{bc}\bar{K}$ & $\Omega^{*}_{bc}\eta$ & $\Xi^{*}_b D$ &  $\Xi^{*}_c \bar{B}$  \\
\hline
    $\Xi^{*}_{bc}\bar{K}$        & $2$  & $\frac{2\sqrt{2}}{\sqrt{3}}$ & $\lambda$    & $0$    \\
    $\Omega^{*}_{bc}\eta$    &         & $0$ & $\frac{\sqrt{2}}{\sqrt{3}}\lambda$    & $0$    \\
    $\Xi^{*}_b D$                    &        &         & $2$                                                   & $0$   \\
    $\Xi^{*}_c \bar{B}$           &        &          &                                                         &    $2$   \\
\hline
\hline
\end{tabular}
\end{table}

\begin{table}[H]
\centering
 \caption{Coefficients $C_{ij}$ for the VB sector with $J^P = \frac{1}{2}^{-}, \frac{3}{2}^{-},  \frac{5}{2}^{-}$.}
 \label{coeff_bc_VB1}
\setlength{\tabcolsep}{6.5pt}
\begin{tabular}{l|cccc}
\hline
\hline
~   &   $\Omega^{*}_{bc}\omega$   & $\Xi^{*}_{bc}\bar{K}^{*}$     & $\Xi^{*}_b D^{*}$     & $\Xi^{*}_c\bar{B}^{*}$ \\
\hline
    $\Omega^{*}_{bc}\omega$         &$0$             & $1$          &  $\lambda$    & $0$  \\ 
    $\Xi^{*}_{bc}\bar{K}^{*}$             &                   & $2$           & $\lambda$    & $0$  \\ 
    $\Xi^{*}_b D^{*}$ 			&  			&  	  	 &$2$ 		 & $0$  \\
    $\Xi^{*}_c\bar{B}^{*}$			& 			& 		 &  			 & $2$  \\
\hline
\hline
\end{tabular}
\end{table}

\subsection{Scattering matrix and pole} 

Once the $V_{ij}$ potential has been calculated, we obtain the scattering matrix in the Bethe-Salpeter equation in coupled channels
\begin{eqnarray}
   T = [1-VG]^{-1}V
\end{eqnarray}
in the matrix form, where $G$ is the diagonal loop function for the meson baryon intermediate state, which we take in the cutoff  
form as in~\cite{Debastiani:2017ewu} with $q_{max}=650$ MeV.

The poles are reached in the second Riemann sheet for which we change $G\to G^{II}$ as
\begin{eqnarray}
G^{II}_j = G^I_j + i \frac{2M_j\,q}{4\pi\sqrt{s}}\,,
\end{eqnarray}
for Re$\sqrt s>m_j+M_j$, and $q$ given by
\begin{eqnarray}
q = \frac{\lambda^{1/2}(s,m^2_j,M^2_j)}{2\sqrt{s}}\,,
\end{eqnarray}
with $m_j$ and $M_j$ the masses of the meson and baryon, respectively. We also evaluate the couplings defined from the residue 
at the pole where the amplitudes go as 
\begin{equation}
T_{ij} = \frac{g_i g_j}{z-z_R} \, ,
\end{equation}
with $z_R$ the complex energy ($M, i\Gamma/2$). We choose one sign for one $g_i$ and the rest of the couplings have the 
relative sign well defined. We also show $g_iG^{II}_i$, which gives the wave function at the origin in coordinate 
space~\cite{Gamermann:2009uq}.

\section{Results} 

In the first place we write in Table~\ref{masses_M_B}, the masses of the mesons and baryons which are needed for the calculations. Those not in the PDG~\cite{ParticleDataGroup:2020ssz} are taken from~\cite{Zhou:2018bkn}. For further discussion about the energies obtained, we also show the masses of the thresholds of the different channels in Table~\ref{Omegacc_1}.

\begin{table}[H]
\centering
\caption{Masses of mesons and baryons in the units of MeV, the values not in the PDG~\cite{ParticleDataGroup:2020ssz} 
               are taken from~\cite{Zhou:2018bkn}.}
 \label{masses_M_B}
\setlength{\tabcolsep}{8pt}
\begin{tabular}{lccccccc}
\hline
\hline
States     & $\bar{K}$ 	& $\eta$  	           & $D$ 		& $\bar{B}$
                 & $\omega$   &$\bar{K}^{*}$    &  $D^{*}$             \\ 
Masses    & $493$          & $548$              & $1870$      & $5279$            
                 & $780$         & $890$              & $2010$                       \\
\hline 
States     & $\bar{B}^{*}$  & $\Xi_c$      & $\Xi'_c$           & $\Xi^{*}_c$   & $\Xi_b$      & $\Xi'_b$      & $\Xi^{*}_b$ \\
Masses    &  $5325$        & $2468$      & $2578$            & $2646$         & $5797$      & $5935$       & $5952$       \\
\hline 
States     & $\Xi_{cc}$ & $\Xi_{cc}^\ast$ & $\Xi_{bb}$ & $\Xi_{bb}^\ast$ & $\Xi_{bc}$ & $\Xi_{bc}^\prime$ & $\Xi_{bc}^\ast$  \\
Masses    & $3622$     & $3675$             & $10340$      & $10370$           & $6922$      & $6948$                  & $6973$ \\
\hline
States     & $\Omega_{cc}$ & $\Omega_{cc}^\ast$ & $\Omega_{bb}$ & $\Omega_{bb}^\ast$ 
                & $\Omega_{bc}$ & $\Omega_{bc}^\prime$ & $\Omega_{bc}^\ast$  \\
Masses    & $3715$     & $3772$             & $10230$      & $10258$           
                & $7011$      & $7047$             & $7066$ \\
\hline
\hline
\end{tabular}
\end{table}

\subsection{Poles and their coupling constants of $\Omega_{cc}$}

In Tables~\ref{cc_coup_1}-\ref{cc_coup_4}
we write the masses of the states obtained, together with the couplings to each channel and the wave function at the origin.
The calculations have been done using $q_{max}=650$ MeV, which was found suited for the study of the $\Omega_c$ states 
in~\cite{Debastiani:2017ewu}, where three experimental states could be associated with molecular states. We write in bold 
characters the case of the biggest coupling, and wave function at the origin, which indicates the most relevant channel.

\begin{table}[H]
\centering
\caption{The poles for $\Omega_{cc}$ along with their coupling constants (in units of MeV) to various channels in the 
              $J^P=\frac{1}{2}^{-}$ sector from $PB(\frac12^+)$.}\label{cc_coup_1}
\setlength{\tabcolsep}{8pt}
\begin{tabular}{l|lcccc}
\hline \hline
    Poles           &       ~           & $\Xi_{cc}\bar{K}$ & $\Omega_{cc}\eta$ & $\Xi_c D$ & $\Xi'_c D$      \\
\hline
  \multirow{2}*{$4069.86$} 
                        &     $g_i$              & $\bm{2.63}$              & $1.55$               & $-1.10$            & $0.26$          \\
            ~          &    $g_iG^{II}_i$   & $\bm{-40.42}$           & $-13.26$            & $3.59$             & $-0.65$         \\
\hline
  \multirow{2}*{$4205.22 +i0.94$} 
                        & $g_i$ 			& $0.10 +i0.20$ 	 & $0.04 +i0.09$ 	& $\bm{6.25-i0.04}$ 	& $0.09+i0.01$     \\
            ~          & $g_iG^{II}_i$ 	& $-5.86-i1.84$  	& $-0.57-i1.32$ 	& $\bm{-31.79+i0.06}$ 	& $-0.30-i0.05$   \\
\hline
  \multirow{2}*{$4310.76 + i0.28$}
                        & $g_i$			 & $0.02+i0.01$ 	&  $-0.13-i0.04$ 	& $-0.02+i0.00$  	& $\bm{6.35+i0.00}$   \\
            ~          & $g_iG^{II}_i$ 	&  $-0.45+i0.64$ 	& $3.47-i0.96$ 	& $0.23-i0.01$ 	& $\bm{-31.95-i0.05}$   \\
     
\hline
\hline
\end{tabular}
\end{table}

\begin{table}[H]
\centering
\caption{The poles for $\Omega_{cc}$ along with their coupling constants (in units of MeV) to various channels in the 
              $J^P=\frac{1}{2}^{-}, \frac{3}{2}^{-}$ sector from $VB(\frac12^+)$.}\label{cc_coup_2}
\setlength{\tabcolsep}{8pt}
\begin{tabular}{l|lcccc}
\hline \hline
    Poles           &       ~           &$\Xi_c D^{*}$ & $\Omega_{cc}\omega$ & $\Xi_{cc}\bar{K}^{*}$  & $\Xi'_c D^{*}$ \\
\hline
  \multirow{2}*{$4332.86$} 
                        & $g_i$ 			&    $\bm{6.51}$   	& $-0.70$		 & $-1.35$  	& $-0.07$ \\
            ~          & $g_iG^{II}_i$ 	&   $\bm{-29.78}$ 	& $5.66$  	& $9.74$ 		& 0.23    	\\
\hline
  \multirow{2}*{$4405.47$} 
                        & $g_i$ 			&$1.27$  		& $1.41$ 		& $\bm{3.81}$   	& $0.83$ 	  \\
            ~          & $g_iG^{II}_i$ 	& $-8.44$ 	& $-15.17$ 	& $\bm{-35.89}$ 	& $-3.33$   \\
\hline
  \multirow{2}*{$4446.29$}
                        & $g_i$			 & $-0.08$ 	 & $-0.32$ 	& $-0.24$  	& $\bm{6.58}$  	\\
            ~          & $g_iG^{II}_i$ 	& $0.73$  	& $4.34$   	& $2.81$  	& $\bm{-30.80}$   	\\
     
\hline
\hline
\end{tabular}
\end{table}

\begin{table}[H]
\centering
\caption{The poles for $\Omega_{cc}$ along with their coupling constants (in units of MeV) to various channels in the 
              $J^P=\frac{3}{2}^{-}$ sector from $PB(\frac32^+)$.}\label{cc_coup_3}
\setlength{\tabcolsep}{8pt}
\begin{tabular}{l|lccc}
\hline \hline
    Poles           &       ~           &$\Xi^{*}_{cc}\bar{K}$ & $\Omega^{*}_{cc}\eta$ & $\Xi^{*}_c D$  \\
\hline
  \multirow{2}*{$4123.85$} 
                        & $g_i$ 			&    $\bm{2.62}$  		 & $1.55$			 & $0.84$ 	\\
            ~          & $g_iG^{II}_i$ 	&  $\bm{-40.61}$		 & $-13.14$ 		&  $-2.09$ 	\\
\hline
  \multirow{2}*{$4380.36 + i0.73$} 
                        & $g_i$ 			&$-0.01-i0.15$		 & $0.02-i0.05$ 	& $\bm{6.28-i0.03}$   	\\
            ~          & $g_iG^{II}_i$ 	& $4.71 +i0.76$	  & $0.41+i1.37$	 & $\bm{-31.94+i0.05}$  	\\     
\hline
\hline
\end{tabular}
\end{table}

\begin{table}[H]
\centering
\caption{The poles for $\Omega_{cc}$ along with their coupling constants (in units of MeV) to various channels in the 
              $J^P=\frac{1}{2}^{-}, \frac{3}{2}^{-}, \frac{5}{2}^{-}$ sector from $VB(\frac32^+)$.}\label{cc_coup_4}
\setlength{\tabcolsep}{8pt}
\begin{tabular}{l|lccc}
\hline \hline
    Poles           &       ~           &$\Omega^{*}_{cc}\omega$  & $\Xi^{*}_{cc}\bar{K}^{*}$ & $\Xi^{*}_c D^{*}$  \\
\hline
  \multirow{2}*{$4446.59$} 
                        & $g_i$ 			& $1.59$ 		& $\bm{3.93}$ 	& $2.64 $ 	\\
            ~          & $g_iG^{II}_i$ 	&$-16.03$ 	& $\bm{-35.31}$ 	& $-9.69$		\\
\hline
  \multirow{2}*{$4520.38$} 
                        & $g_i$ 			&$-0.18$ 		& $-0.94$	 	& $\bm{6.10}$ 		\\
            ~          & $g_iG^{II}_i$ 	& $2.78 $ 	& $12.44 $ 	& $\bm{-29.41}$  	\\     
\hline
\hline
\end{tabular}
\end{table}

By looking at Table~\ref{cc_coup_1} we observe that we obtain states at $4070, 4205, 4311$~MeV. The widths, corresponding to twice the imaginary part at the pole, are all below $2$~MeV, the most relevant channels are 
the $\Xi_{cc}\bar{K}$ for the $4070$~MeV, the $\Xi_{c}D$ for the $4205$~MeV and the  $\Xi'_{c}D$ for the $4311$~MeV. 
The $\Omega_{cc}\eta$ channel has a relatively important weight also in the $4070$ MeV state, but is negligible in the other states.
By looking at the thresholds in the Table~\ref{Omegacc_1}, we see that the $4070$~MeV state could mostly qualify as a $\Xi_{cc}\bar{K}$ molecule with a binding of about $45$~MeV, the $4311$ state would be mostly a $\Xi_{c}D$ state bound by about $133$~MeV and the $4311$ would correspond to a $\Xi'_{c}D$ state bound by about $137$~MeV. However we should not ignore that we have a mixture of coupled channels in the wave functions and some components are less bound than others, hence it is not fully appropriate to put all the binding in just one component, the total energies being the relevant magnitudes to be considered. In Table~\ref{cc_coup_2}  we find similar features to  the former one with three states that couple mostly to $\Xi_{c}D^*$, $\Xi_{cc}\bar{K}^*$, and $\Xi'_{c}D^*$ respectively.

In Table~\ref{cc_coup_3} we obtain two states of $PB(\frac{3}{2}^+)$ nature at $4124$~MeV and $4380$~MeV, which couple mostly to $\Xi^*_{cc}\bar{K}$ and $\Xi^*_{c}D$ respectively. The widths are also smaller than $2$~MeV. In Table~\ref{cc_coup_4}  we also obtain two states of $VB(\frac{3}{2}^+)$ nature, and hence $J^P=\frac{1}{2}^-, \frac{3}{2}^-, \frac{5}{2}^-$, that couple mostly to  $\Xi^*_{cc}\bar{K}^*$ and $\Xi^*_{c}D^*$ respectively. The widths are null with the space of states considered, hence we expect them to be very small.

\subsection{Poles and their coupling constants of $\Omega_{bb}$}

In Table~\ref{Omegabb_1} we put the threshold of the channels involved in the calculations.
In Tables~\ref{bb_coup_1} - \ref{bb_coup_4}, we show the bound states and resonances of $\Omega_{bb}$ as well as their 
coupling constants to various channels, obtained with $q_{max}=650$ MeV.
\begin{table}[H]
\centering
\caption{The poles for $\Omega_{bb}$ along with their coupling constants (in units of MeV) to various channels in the 
              $J^P=\frac{1}{2}^{-}$ sector from $PB(\frac12^+)$.}\label{bb_coup_1} 
\setlength{\tabcolsep}{8pt}
\begin{tabular}{l|lcccc}
\hline \hline
    Poles           &       ~           & $\Omega_{bb}\eta$ & $\Xi_{bb}\bar{K}$ &  $\Xi_b \bar{B}$&  $\Xi'_b \bar{B}$        \\
\hline
  \multirow{2}*{$10741.65$} 
                        &     $g_i$              & $1.50$   		& $\bm{2.72}$ 		& $0$ 		& $0$               \\
            ~          &    $g_iG^{II}_i$   & $-25.56$ 	& $\bm{-34.78}$ 	& $0$ 		& $0$       	 \\
\hline
  \multirow{2}*{$10864.15$} 
                        &     $g_i$              & $0$   		& $0$ 		& $\bm{11.87}$ 	& $0$        	  \\
            ~          &    $g_iG^{II}_i$   & $0$ 		& $0$  		& $\bm{-20.43}$ 	& $0$      		  \\
\hline
  \multirow{2}*{$11001.63$} 
                        &     $g_i$              & $0$   		& $0$ 		& $0$ 		& $\bm{11.87}$      	    \\
            ~          &    $g_iG^{II}_i$   & $0$   		& $0$ 		& $0$  		& $\bm{-20.43}$     	   \\     
\hline
\hline
\end{tabular}
\end{table}

\begin{table}[H]
\centering
\caption{The poles for $\Omega_{bb}$ along with their coupling constants (in units of MeV) to various channels in the 
              $J^P=\frac{1}{2}^{-}, \frac{3}{2}^{-}$ sector from $VB(\frac12^+)$.}\label{bb_coup_2} 
\setlength{\tabcolsep}{8pt}
\begin{tabular}{l|lcccc}
\hline \hline
    Poles           &       ~           &$\Omega_{bb}\omega$  & $\Xi_b \bar{B}^{*}$ & $\Xi_{bb}\bar{K}^{*}$ & $\Xi'_b \bar{B}^{*}$  \\
\hline
  \multirow{2}*{$10909.88$} 
                        & $g_i$ 			& $0$  &$\bm{11.92}$ & $0$ & $0$ \\
            ~          & $g_iG^{II}_i$ 	& $0$  & $\bm{-20.35}$ & $0$ & $0$    	\\
\hline
  \multirow{2}*{$11047.36$} 
                        & $g_i$ 			& $0$   & $0$ & $0$ &$\bm{11.92}$ \\
            ~          & $g_iG^{II}_i$ 	& $0$  & $0$ & $0$    & $\bm{-20.34}$  	\\
\hline
\hline
\end{tabular}
\end{table}

\begin{table}[H]
\centering
\caption{The poles for $\Omega_{bb}$ along with their coupling constants (in units of MeV) to various channels in the 
              $J^P=\frac{3}{2}^{-}$ sector from $PB(\frac32^+)$.}\label{bb_coup_3} 
\setlength{\tabcolsep}{8pt}
\begin{tabular}{l|lccc}
\hline \hline
    Poles           &       ~           & $\Omega^{*}_{bb}\eta$ & $\Xi^{*}_{bb}\bar{K}$  & $\Xi^{*}_b \bar{B}$  \\
\hline
  \multirow{2}*{$10770.91$} 
                        & $g_i$ 			&   $1.50 $   & $\bm{2.71}$ & $0$	\\
            ~          & $g_iG^{II}_i$ 	&  $-25.70$ & $\bm{-34.62}$ &  $0$ 	\\
\hline
  \multirow{2}*{$11018.56$} 
                        & $g_i$ 			&$0 $ & $0$   &$\bm{11.87}$	\\
            ~          & $g_iG^{II}_i$ 	& $0 $ & $0$  & $\bm{-20.43}$ 	\\     
\hline
\hline
\end{tabular}
\end{table}

\begin{table}[H]
\centering
\caption{The poles for $\Omega_{bb}$ along with their coupling constants (in units of MeV) to various channels in the 
              $J^P=\frac{1}{2}^{-}, \frac{3}{2}^{-}, \frac{5}{2}^{-}$ sector from $VB(\frac32^+)$.}\label{bb_coup_4} 
\setlength{\tabcolsep}{8pt}
\begin{tabular}{l|lccc}
\hline \hline
    Poles           &       ~           & $\Omega^{*}_{bb}\omega$ & $\Xi^{*}_{bb}\bar{K}^*$  & $\Xi^{*}_b \bar{B}^*$  \\
\hline
  \multirow{2}*{$11064.30$} 
                        & $g_i$ 			& $0 $ & $0$   & $\bm{11.94}$	\\
            ~          & $g_iG^{II}_i$ 	& $0 $ & $0$  & $\bm{-20.37}$ 	\\     
\hline
\hline
\end{tabular}
\end{table}
In Table~\ref{bb_coup_1} we show the states of $J^P=\frac{1}{2}^{-}$ coming from $PB(\frac{1}{2}^{+})$. We obtain three states, 
one at $10742$~MeV, another one at $10864$~MeV and another one at $11002$~MeV, coupling mostly to $\Xi_{bb}\bar{K}$, 
$\Xi_{b}\bar{B}$ and $\Xi'_{b}\bar{B}$, respectively. The $VB(\frac{1}{2}^{+})$ channels shown in Table~\ref{bb_coup_2}
give rise to one state at $10910$~MeV coupling to $\Xi_{b}\bar{B}^*$ and another one at  $11047$~MeV coupling to 
$\Xi^\prime_{b}\bar{B}^*$.  The widths in this case are null with the channels chosen and the approximations done.

For the case $PB(\frac{3}{2}^{+})$ states, shown in Table~\ref{bb_coup_3}, we obtain two states with $J^P=\frac{3}{2}^{-}$, 
coupling mostly to  $\Xi^*_{bb}\bar{K}$ and $\Xi^*_{b}\bar{B}$ respectively. 
Finally, in Table~\ref{bb_coup_4} we show the only state that we get for  $VB(\frac32^+)$, which couples to $\Xi^{*}_b \bar{B}^*$.

\subsection{Poles and their coupling constants of $\Omega_{bc}$}

We put the results for the threshold masses in Table~\ref{Omegabc_1}. 
In Tables~\ref{bc_coup_1} - \ref{bc_coup_4}, we show the bound states and resonances of $\Omega_{bc}$ along with their 
coupling constants to various channels with $q_{max}=650$ MeV.

\begin{table}[H]
\footnotesize
\centering
\caption{The poles for $\Omega_{bc}$ along with their coupling constants (in units of MeV) to various channels in the 
              $J^P=\frac{1}{2}^{-}$ sector from $PB(\frac12^+)$.}\label{bc_coup_1}
\setlength{\tabcolsep}{2pt}
\begin{tabular}{l|lcccccccc}
\hline \hline
    Poles           &       ~           & $\Xi_{bc}\bar{K}$ & $\Xi'_{bc}\bar{K}$  & $\Omega_{bc}\eta$ & $\Omega'_{bc}\eta$  
                                              & $\Xi_b D$             & $\Xi_c \bar{B}$       & $\Xi'_b D$                & $\Xi'_c \bar{B}$     \\
\hline
  \multirow{2}*{$7362.26$} 
                        &     $g_i$              &$\bm{2.64}$ 	& $0$ 	& $1.57$ 	  & $0$ 
                                                      &$1.70$ 	& $0$  	& $0$  	  & $0$           \\
            ~          &    $g_iG^{II}_i$   &$\bm{-40.41}$ & $0$       & $-13.52$ & $0$   
                                                      & $-5.35$  & $0$       & $0$          & $0$     	 \\
\hline
  \multirow{2}*{$7392.60$} 
                        &     $g_i$              & $0$  		& $\bm{2.61}$ 		& $0$ 	  	& $1.51$ 
                                                      & $0$  		& $0$  		& $-0.73$  	  	& $0$           	\\
            ~          &    $g_iG^{II}_i$   & $0$ 		& $\bm{-41.08}$       & $0$ 	  	& $-12.83$   
                                                      & $0$  		& $0$       	& $1.81$         	& $0$       	 \\
\hline
  \multirow{2}*{$7514.32 + i2.21$} 
                        &     $g_i$              & $-0.14-i0.27$  & $0$  		& $-0.05-i0.13$	  	& $0$ 
                                                      & $\bm{6.19 -i0.08}$  & $0$  		& $0$  	  			& $0$           	\\
            ~          &    $g_iG^{II}_i$   & $9.18 +i2.42$   & $0$       	& $0.83 +i2.04$ 	  	& $0$   
                                                      & $\bm{-32.11+i0.12}$ & $0$       	& $0$         			& $0$       	 \\                
\hline
  \multirow{2}*{$7566.65$} 
                        &     $g_i$              & $0$  		& $0$ 			& $0$ 	  	& $0$ 
                                                      & $0$  		& $\bm{11.50}$  		& $0$  	  	& $0$           	\\
            ~          &    $g_iG^{II}_i$   & $0$ 		& $0$       		& $0$ 	  	& $0$   
                                                      & $0$  		& $\bm{-20.01}$       	& $0$         	& $0$       	 \\         
\hline
  \multirow{2}*{$7641.20+i 2.26$} 
                        &     $g_i$              & $0$  		& $-0.06-i0.03$ 	& $0$ 	  		& $0.34+i0.11$ 
                                                      & $0$  		& $0$  			& $\bm{6.50+i0.02}$  	& $0$           	\\
            ~          &    $g_iG^{II}_i$   & $0$ 		& $1.60-i1.76$       	& $0$ 	  		& $-10.29+i2.74$   
                                                      & $0$  		& $0$       	        & $-\bm{32.20-i0.41}$     & $0$       	 \\             
\hline
  \multirow{2}*{$7674.29$} 
                        &     $g_i$              & $0$  		& $0$ 		& $0$ 	  	& $0$ 
                                                      & $0$  		& $0$  		& $0$  	  	& $\bm{11.53}$           	\\
            ~          &    $g_iG^{II}_i$   & $0$ 		& $0$       	& $0$ 	  	& $0$   
                                                      & $0$  		& $0$       	& $0$         	& $\bm{-20.05}$      	 \\                                                                                                                                                                                                              
\hline
\hline
\end{tabular}
\end{table}

\begin{table}[H]
\centering
\caption{The poles for $\Omega_{bc}$ along with their coupling constants (in units of MeV) to various channels in the 
               $J^P=\frac{1}{2}^{-}, \frac{3}{2}^{-}$ sector  from $VB(\frac12^+)$.}\label{bc_coup_2}
\setlength{\tabcolsep}{8pt}
\begin{tabular}{l|lcccccccc}
\hline \hline
    Poles           &       ~           & $\Omega_{bc}\omega$ & $\Xi_c \bar{B}^{*}$ & $\Xi_b D^{*}$  & $\Xi_{bc}\bar{K}^{*}$ 
                                               & $\Omega'_{bc}\omega$  & $\Xi'_{bc}\bar{K}^{*}$  &  $\Xi'_c \bar{B}^{*}$  & $\Xi'_b D^{*}$   \\
\hline
  \multirow{2}*{$7612.44$} 
                        &     $g_i$              & $0$  		& $\bm{11.56}$ 		& $0$ 	  	& $0$ 
                                                      & $0$  		& $0$  		& $0$  	  	& $0$           	\\
            ~          &    $g_iG^{II}_i$   & $0$ 		& $\bm{-19.93}$       	& $0$ 	  	& $0$   
                                                      & $0$  		& $0$       	& $0$         	& $0$       	 \\                     
\hline
  \multirow{2}*{$7627.73$} 
                        &     $g_i$              & $1.09$  		& $0$ 		& $\bm{6.36}$ 	  	& $2.14$ 
                                                      & $0$  		& $0$  		& $0$  	  	& $0$           	\\
            ~          &    $g_iG^{II}_i$   & $-9.13$ 		& $0$       	& $\bm{-28.05}$ 	 & $-15.65$   
                                                      & $0$  		& $0$       	& $0$         	& $0$       	 \\                                                                                            
\hline
  \multirow{2}*{$7707.67$} 
                        &     $g_i$              &$1.19$ 	& $0$ 	& $-2.17$ 	  & $\bm{3.40}$ 
                                                      &$0$ 	& $0$  	& $0$  	  & $0$           \\
            ~          &    $g_iG^{II}_i$   &$-13.85$ & $0$       & $14.00$          & $\bm{-33.94}$   
                                                      & $0$  & $0$       & $0$          & $0$     	 \\
\hline
  \multirow{2}*{$7716.28$} 
                        &     $g_i$              & $0$  		& $0$ 		& $0$ 	  	& $0$
                                                      & $1.43$  		& $\bm{4.03}$  	& $0$  	  	& $-1.77$           	\\
            ~          &    $g_iG^{II}_i$   & $0$ 		& $0$       	& $0$ 	  	& $0$   
                                                      & $-14.61$ 	& $\bm{-37.19}$       & $0$         	& $6.54$       	 \\                    
\hline
  \multirow{2}*{$7720.07$} 
                        &     $g_i$              & $0$  		& $0$ 		& $0$ 	  		& $0$ 
                                                      & $0$  		& $0$  		& $\bm{11.59}$  	  	& $0$           	\\
            ~          &    $g_iG^{II}_i$   & $0$ 		& $0$       	& $0$ 	  		& $0$   
                                                      & $0$  		& $0$       	& $\bm{-19.97}$         	& $0$       	 \\                                                                                                                                                                                                                                                                                                                                                      
\hline
  \multirow{2}*{$7777.47$} 
                        &     $g_i$              & $0$  		& $0$ 		& $0$ 	  	& $0$ 
                                                      & $0.78$  		& $0.38$  		& $0$  	  	& $\bm{6.50}$           	\\
            ~          &    $g_iG^{II}_i$   & $0$ 		& $0$       	& $0$ 	  	& $0$   
                                                      & $-11.04$  		& $-4.84$       	& $0$         	& $\bm{-30.09}$       	 \\       
\hline
\hline
\end{tabular}
\end{table}

\begin{table}[H]
\centering
\caption{The poles for $\Omega_{bc}$ along with their coupling constants (in units of MeV) to various channels in the 
              $J^P=\frac{3}{2}^{-}$ sector from $PB(\frac32^+)$.}\label{bc_coup_3}
\setlength{\tabcolsep}{8pt}
\begin{tabular}{l|lcccc}
\hline \hline
    Poles           &       ~           & $\Xi^{*}_{bc}\bar{K}$ & $\Omega^{*}_{bc}\eta$ & $\Xi^{*}_b D$ &   $\Xi^{*}_c \bar{B}$  \\
\hline
  \multirow{2}*{$7415.55$} 
                        & $g_i$ 			& $\bm{2.63}$   	& $1.56$  	&$1.21$  		&  $0$	\\
            ~          & $g_iG^{II}_i$ 	& $\bm{-40.83}$  	& $-13.37$ 	& $-3.05$ 	&  $0$ 	\\
\hline
  \multirow{2}*{$7667.65+i1.40$} 
                        & $g_i$ 			&$-0.02-i0.20$ 	&$0.02-i0.06$   	& $\bm{6.25-i0.05}$ 	&$0$	\\
            ~          & $g_iG^{II}_i$ 	&$6.82+i0.98$   	& $0.53+i1.88$  	& $\bm{-32.26+i0.09}$	& $0$       \\     
\hline
  \multirow{2}*{$7740.93$} 
                        & $g_i$ 			&$0$   	& $0$  	&$0$	& $\bm{11.52}$	\\
            ~          & $g_iG^{II}_i$ 	& $0$  	&$0$ 	& $0$ 	& $\bm{-20.08}$ 	\\
\hline
\hline
\end{tabular}
\end{table}

\begin{table}[H]
\centering
\caption{The poles for $\Omega_{bc}$ along with their coupling constants (in units of MeV) to various channels in the 
              $J^P=\frac{1}{2}^{-}, \frac{3}{2}^{-}, \frac{5}{2}^{-}$ sector from $VB(\frac32^+)$.}\label{bc_coup_4}
\setlength{\tabcolsep}{8pt}
\begin{tabular}{l|lcccc}
\hline \hline
    Poles           &       ~           & $\Omega^{*}_{bc}\omega$  & $\Xi^{*}_{bc}\bar{K}^{*}$ & $\Xi^{*}_b D^{*}$ & $\Xi^{*}_c\bar{B}^{*}$ \\
\hline
  \multirow{2}*{$7729.11$} 
                        & $g_i$ 			& $1.60$  	& $\bm{3.82}$  	& $3.54$  	& $0$ 	\\
            ~          & $g_iG^{II}_i$ 	& $-15.96$ 	& $\bm{-33.56}$ 	& $-12.92$ 	& $0$	\\     
\hline
  \multirow{2}*{$7786.71$} 
                        & $g_i$ 			& $0$  		& $0$  		& $0$  		& $\bm{11.61}$ 	\\
            ~          & $g_iG^{II}_i$ 	& $0$ 		& $0$ 		& $0$ 		& $\bm{-19.99}$	\\   
\hline
  \multirow{2}*{$7811.82$} 
                        & $g_i$ 			& $-0.23$ 	& $-1.24$ 	&  $\bm{5.71}$   	& $0$ 	\\
            ~          & $g_iG^{II}_i$ 	&$3.72$   	& $16.77$ 	& $\bm{-28.48}$  	& $0$ 	\\              
\hline
\hline
\end{tabular}
\end{table}

For the case of $PB(\frac{1}{2}^{+})$ states, shown in Table~\ref{bc_coup_1}, we find six states coupling mostly to 
$\Xi_{bc}\bar{K}$, $\Xi'_{bc}\bar{K}$, $\Xi_b D$, $\Xi_c \bar{B}$, $\Xi'_b D$, and $\Xi'_c \bar{B}$, respectively.
For the case of $VB(\frac{1}{2}^{+})$ we find six states, shown in Table~\ref{bc_coup_2}, 
coupling mostly to $\Xi_c \bar{B}^{*}$, $\Xi_b D^{*}$, $\Xi_{bc}\bar{K}^{*}$, $\Xi'_{bc}\bar{K}^{*}$, $\Xi'_c \bar{B}^{*}$, 
and $\Xi'_b D^{*}$, respectively. 

For the case of $PB(\frac{3}{2}^{+})$ with $J^P=\frac{3}{2}^{-}$ we show the states found in Table~\ref{bc_coup_3}. We obtain three states coupling mostly to  $\Xi^*_{bc}\bar{K}$,  $\Xi^*_{b}D$, and $\Xi^*_{c}\bar{B}$, respectively. 
The widths are also small, all of them below $3$~MeV. For the case of  $VB(\frac{3}{2}^{+})$ we obtain three states of $J^P=\frac{1}{2}^{-}, \frac{3}{2}^-, \frac{5}{2}^-$, shown in Table~\ref{bc_coup_4}, coupling mostly to $\Xi^*_{bc}\bar{K}^*$,  $\Xi^*_{c}\bar{B}^*$, and $\Xi^*_{b}{D}^*$ respectively.  

The widths obtained are small in all cases. In the cases of $VB(\frac{1}{2}^{+})$, $PB(\frac{3}{2}^{+})$, and $VB(\frac{3}{2}^{+})$ there
can be transitions to the $PB(\frac{1}{2}^{+})$ states, but we anticipated that these transitions are very suppressed and the widths should be smaller than those found for transitions allowed by vector exchange within the blocks considered. For the case of 
$PB(\frac{1}{2}^{+})$, the $4070$ MeV state that couples to $\Xi_{cc}\bar{K}$ cannot decay to any other state in our space, since
it is bound in $\Xi_{cc}\bar{K}$ and this channel has the smallest threshold. The state with $4205$ MeV can only decay to 
$\Xi_{cc}\bar{K}$,  so this should be the channel to observe it. The state at $4311$~MeV can decay to $\Omega_{cc}\eta$ and 
$\Xi_{cc}\bar{K}$. Given the couplings to the channels in Table~\ref{cc_coup_1}, the favored channel for observation would be  
$\Omega_{cc}\eta$. Similar considerations can be done in the other sectors.

\section{Estimation of uncertainties}\label{unces}   

As was done in Ref.~\cite{Debastiani:2017ewu}, we estimate uncertainties in the results, as a consequence of uncertainties 
in the cut off, or the strength of the interaction governed by the parameter $f_\pi$. The value of $q_{max}$ has been taken 
as $650$ MeV, as in~\cite{Debastiani:2017ewu}, where three experimental $\Omega_c$ states were well reproduced. 
Given the analogy of the states studied here, this should be a good starting point to have a realistic estimate of the states that 
appear from the interaction in coupled channels. However, it is also interesting to see what uncertainties we can have in the masses, 
widths and couplings obtained. For this we follow the same strategy as in Ref.~\cite{Debastiani:2017ewu} and repeat the calculations 
for $q_{max}=600,~700$ MeV or $f_\pi=97.6$ MeV (rather than $93$ MeV). We also perform another sort of calculation. 
We position ourselves in the hypothetical situation where the experiment has been done and the states predicted are observed. 
A qualitative agreement with the experimental results is expected, but, as usually done, we would take one experimental mass 
and do fine tuning of $q_{max}$ to get this mass. The rest of masses and widths would then be genuine predictions.  

The exercise has been done for the $\Omega_{bc}$ states, where more states appear. We have checked that in the other sectors, the 
conclusions are the same, and for the sake of conciseness we report the results in the $\Omega_{bc}$ sector. The results are shown 
in Table~\ref{bc_uncert}.

We observe in Table~\ref{bc_uncert} that a change of $50$ MeV in $q_{max}$ reverts in changes of the masses by $10$-$35$ MeV. 
In the case of the $3/2^+$ states the difference can even be bigger, up to $50$ MeV. The change of $f_\pi$ keeping $q_{max}=650$ 
MeV induces changes of $10$-$20$ MeV in the mass.

More interesting is to see the change when we change $f_\pi$ and at the same time $q_{max}$ in order to have the same mass of 
one of the states, in this case the $PB(1/2^+)$ state at $7362.26$ MeV. There we find that the other masses change within the 
range $0$-$10$ MeV and the widths change within $10\%$. The couplings to the main channel also change within $10\%$. This 
tells us the level of accuracy that we can expect once the experiments are performed, and some states are observed, and we fine tune 
our parameters to obtain the mass of one of the states.

\begin{table}[H]
\footnotesize
\centering
\caption{The poles for $\Omega_{bc}$ with different $q_{max}$ and $f_\pi$. }\label{bc_uncert}
\setlength{\tabcolsep}{12pt}
\begin{tabular}{l|ccccc}
\hline \hline
  \multirow{2}*{~} 
                        & $f_\pi=93$              & $f_\pi=93$            & $f_\pi=93$              & $f_\pi=97.6$             & $f_\pi=97.6$      \\
            ~         & $q_{max}=600$       & $q_{max}=650$   & $q_{max}=700$      & $q_{max}=650$        & $q_{max}=692.2$       \\                        
\hline
  \multirow{6}*{$PB(\frac12^+)$} 
                        &     $7377.58$  &$7362.26$ 	& $7345.03$ 	 & $7374.90$ 	  & $7362.26$         \\
            ~          &    $7406.39$   &$7392.60$ & $7377.65$       & $7404.44$  & $7393.32$   	 \\
            ~          &    $7547.32+i1.72$   &$7514.32+i2.21$ & $7477.58+i2.58$       & $7532.81+i2.16$ & $7504.85+i2.59$   	 \\
            ~          &    $7606.82$   &$7566.65$ & $7520.57$       & $7587.93$ & $7553.26$   	 \\
            ~          &    $7676.65+i2.78$   &$7641.20+i2.26$ & $7601.40+i0.95$       & $7660.99+i2.52$ & $7630.80+i2.00$   	 \\
            ~          &    $7714.92$   &$7674.29$ & $7627.67$       & $7695.67$ & $7660.58$   	 \\
\hline
  \multirow{6}*{$VB(\frac12^+)$} 
                        &    $7652.65$   &$7612.44$ & $7566.30$       & $7633.72$ & $7599.01$   	 \\
            ~          &    $7665.15$   &$7627.73$ & $7586.28$       & $7648.39$ & $7616.67$   	 \\
            ~          &    $7730.21$   &$7707.67$ & $7682.99$       & $7722.76$ & $7704.10$   	 \\
            ~          &    $7742.10$   &$7716.28$ & $7687.80$       & $7733.27$ & $7711.72$   	 \\
            ~          &    $7760.74$   &$7720.07$ & $7673.40$       & $7741.46$ & $7706.33$   	 \\
            ~          &    $7811.99$   &$7777.47$ & $7738.99$       & $7796.55$ & $7767.20$   	 \\                                                 
\hline
  \multirow{3}*{$PB(\frac32^+)$} 
                        &    $7429.99$   &$7415.55$ & $7399.73$       & $7427.74$ & $7416.00$   	 \\
            ~          &    $7701.15+i0.95$   &$7667.65+i1.40$ & $7630.12+i1.83$       & $7686.35+i1.30$ & $7657.87+i1.67$   	 \\
            ~          &    $7781.82$   &$7740.93$ & $7693.99$       & $7762.36$ & $7727.03$   	 \\
\hline
  \multirow{3}*{$VB(\frac32^+)$} 
                        &    $7758.72$   &$7729.11$ & $7695.70$       & $7747.68$ & $7722.48$   	 \\
            ~          &    $7827.64$   &$7786.71$ & $7739.72$       & $7808.15$ & $7772.78$   	 \\
            ~          &    $7842.05$   &$7811.82$ & $7779.12$       & $7828.91$ & $7803.79$   	 \\                                                                                                                                                                                            
\hline
\hline
\end{tabular}
\end{table}

\section{Conclusion}

We have done a thorough study of the molecular states of type $\Omega_{cc}, \Omega_{bb}, \Omega_{bc}$ that stem from the interaction of meson baryon coupled channels with these quantum numbers. We classify them as $PB(\frac{1}{2}^{+})$, $PB(\frac{3}{2}^{+})$, $VB(\frac{1}{2}^{+})$,  $VB(\frac{3}{2}^{+})$, hence, channels composed of a meson, pseudoscalar or vector, and a baryon in its ground state with either spin $\frac{1}{2}$ or $\frac{3}{2}$. The interaction is evaluated using an extension of the local hidden gauge approach and we only consider $S$-wave. Hence we obtain states carrying $J^P=\frac{1}{2}^{-}, \frac{3}{2}^-, \frac{5}{2}^-$. In the case of $VB(\frac{1}{2}^{+})$ we have degenerate states in $J^P=\frac{1}{2}^{-}, \frac{3}{2}^-$ and in the case of $VB(\frac{3}{2}^{+})$ we obtain degenerate states with $J^P=\frac{1}{2}^{-}, \frac{3}{2}^-, \frac{5}{2}^-$. We obtain states of each type for the three 
$\Omega_{cc}, \Omega_{bb}, \Omega_{bc}$ sectors. We look for poles of the scattering matrix in the second Riemann sheet and then evaluate the couplings of the states obtained to each channel. Simultaneously, we also evaluate the wave function at the origin. In all the states observed we find one channel that has a much bigger coupling and wave function at the origin than the other channels, which we identify as the main component of the wave function of that state in terms of the coupled channels considered. Although in the case of coupled channels it is difficult to define a binding, if we refer to the threshold of the main component, we find bindings of the order of $50$-$130$~MeV. These bindings are in the line of bindings obtained for other case as  $\Omega_{c}, \Xi_{c}, \Xi_{b}$ etc., whose agreement with some states found  experimentally has been reported.

The states that we have chosen to study here are presently under analysis by the LHCb collaboration and it will be most instructive to compare with the experimental results whenever they are available.

\section{acknowledgement}
W. F. Wang and J. Song thank Meng-Lin Du for valuable discussions. J. Song also wishes to acknowledge support from China Scholarship Council. This work was supported in part by the National Natural Science Foundation of China under Grant No. 12147215. This work was partly supported by the Spanish Ministerio de Economia y Competitividad (MINECO) and 16 European FEDER funds under Contracts No. PID2020-112777GB-I00, and by Generalitat Valenciana under contract PROMETEO/2020/023. This project has received funding from the European Union Horizon 2020 research and innovation programme under the program H2020-INFRAIA-2018-1, grant agreement No. 824093 of the STRONG-2020 project. The work of A. F. was partially supported by the Generalitat Valenciana and European Social Fund APOSTD-2021-112.


\end{document}